\documentstyle[aps,prl,floats,epsfig]{revtex}	

\title{Hall magnetocapacitance in two-dimensional electron systems}
\author{A.M.C. Valkering$^{1}$, P.K.H. Sommerfeld\cite{present}$^{1}$, R.A.M. van de Ven$^{1}$, R.W. van der Heijden$^{1}$, F.A.P. Blom$^{1}$, M.J. Lea$^{2}$ and F.M. Peeters$^{3}$}
\address{$^{1}$COBRA Interuniversity Research Institute, Department of Physics, 
Eindhoven
University of Technology, P.O. Box 513, NL-5600
MB Eindhoven, The Netherlands}
\address{$^{2}$Department of Physics, Royal Holloway, University of London, Egham, 
TW20 0EX, UK}
\address{$^{3}$Department of Physics, University of Antwerp, Universiteitsplein 1, B-2610 Antwerpen, Belgium} 
\date{\today}

\begin{document}
\draft

\twocolumn[\hsize\textwidth\columnwidth\hsize\csname@twocolumnfalse\endcsname

\maketitle

\begin{abstract}
The magnetocapacitance of a two-dimensional electron system (2DES) is
investigated experimentally, both under and away from Quantum Hall (QH) conditions, at frequencies between 1 kHz and 100 MHz.
The nature of the capacitive signal in a bounded 2DES is determined by a resistive cut-off frequency 1/$\tau \propto \sigma_{xx}$, the longitudinal magnetoconductivity. A new response mechanism is reported for angular frequencies $\omega > 1/\tau$, which is controlled by the $\em transverse$ or Hall conductivity $\sigma_{xy}$ and the boundaries of the sample, even at frequencies far below those of the Edge Magnetoplasma resonances and away from the QH-conditions.
\end{abstract}


\vskip2pc]

\narrowtext

\newcommand{\av}[1]{\mbox{$\langle #1 \rangle$}}

The spatial dependence of the electrical properties, and in particular the role of the sample-edges, remains one of the major issues of present-day research of two-dimensional electron systems (2DES) in the Quantum
Hall (QH) state \cite{thouless}. A large number of experiments, employing finite-frequency methods, has recently been reported on qualitative and quantitative investigations on the current or charge distributions. Inductive techniques have been used to investigate Hall currents in the bulk of the sample \cite{yahel}. 
Magnetocapacitance data have yielded evidence for quantum edge channels \cite{taka}, but on the other hand display properties that are clearly of classical origin though they would be more readily understood in terms of quantum channels \cite{chen,sommerfeld}. Such techniques have also been applied to test theories for composite fermions in the fractional QH effect regime \cite{moon} or to study as yet poorly understood phenomena associated with in-plane tunneling of electrons \cite{zhitenev}. Of particular importance are direct imaging techniques of various kind \cite{fontein,dietsche}. Very recently, novel high-resolution scanning probe imaging techniques have been developed, some of which depend basically on AC-techniques \cite{tessmer,today}. Low-frequency (100 kHz) AC measurements give quite different results from DC methods and are the key to imaging mobile charges \cite{today}.  

Despite its increasing use, a thorough understanding of magnetocapacitance or more generally finite-frequency properties, is 
still lacking. Traditionally, (magneto)capacitance is used to obtain information 
on the density of states (DOS) in the 2DES 
\cite{kaplit}. Since it became clear that the longitudinal conductivity 
$\sigma_{xx}$ strongly affects the measured magnetocapacitance \cite{good}, 
it is now frequently used to study transport properties. Most previous experiments measured the signal between two circularly symmetric and capacitively-coupled electrodes in a Corbino geometry. No signal is then obtained if $\sigma_{xx}=0$, as in the low temperature QH region \cite{good}. Intuitively, $\sigma_{xx}$ is expected to be essential for capacitive coupling and consequently it has been suggested that "in the true Quantum Hall region, it is impossible to couple capacitively to a 2DE[S], since no charge flow can occur" \cite{stiles}. 

Dynamical effects in strong magnetic fields for more general arrangements than 
the Corbino, at frequencies much lower than any characteristic frequency of the 
2DES were investigated  by Lea {\it et al.} \cite{lea} for the 
classical system of electrons on liquid helium, fully screened by metallic plates parallel to the 2DES. Independently, they were studied for the screened semiconductor 2DES by Grodnensky {\it et al.} 
\cite{grodnensky}. In the experimental arrangement of 
Ref. \cite{grodnensky} the magnetocapacitance varied with magnetic field by only 
a few percent, which could suggest that the Hall coupling is of marginal importance only. Moreover, the distinct roles of {\em both} tensor components $\sigma_{xx}$ and $\sigma_{xy}$ were not explicitly revealed. In addition, from the results of \cite{grodnensky}, it was not clear what the implications are for more conventional, low-frequency capacitance measurements (e.g. Ref. \cite{taka}). 

In the present work, we demonstrate experimentally that a capacitive coupling can also arise because of charge accumulation at the edge due to the Hall effect. The charge can propagate along the edge as a result of the transverse conductivity $\sigma_{xy}$. A direct experimental proof is obtained by the use of a grounded Ohmic contact in the centre of a square sample, which eliminates any conduction across the bulk of the 2DES due to finite $\sigma_{xx}$. It is shown that this new coupling mechanism is effective irrespective of the Quantum Hall conditions. The relative importance of the $\sigma_{xx}$ and $\sigma_{xy}$ contributions is controlled by an $RC$ time constant $\tau = \varepsilon_{0} \varepsilon_{r} W/\sigma_{xx}$ \cite{grodnensky}, which is the time it takes excess charge to diffuse out over the entire sample with lateral dimension $W$. For angular frequencies $\omega > 1/\tau$, the $\sigma_{xx}$-contribution is ineffective. The condition $\omega \tau > 1$ is easily fulfilled at the QH-regions because of the vanishing $\sigma_{xx}$. Outside the QH-regions it is satisfied in the present work by using sufficiently high frequencies $\omega$. Our results predict that a finite residual capacitance will occur for $\omega \tau \gg 1$ both at and outside the QH-regions, for the real, smooth density profile of the 2DES near the sample boundary. No explicit appeal to metallic edge channels is required. 

The samples are obtained from a modulation-doped GaAs-AlGaAs heterostructure 
with 2DES density 
$n=1.6 \times 10^{15}$~m$^{-2}$ and mobility $\mu =$80~m$^{2}$/Vs. Two 10$\times$10 mm$^{2}$ 
samples were cleaved from this wafer. One was not further treated and coupled 
along opposite sides to 
two pieces of centre conductor wire out of a coaxial cable, which serve as excitation and detection electrodes similar as in 
Ref.~\cite{grodnensky} (see inset in Fig. 1). A 
7$\times$7~mm$^{2}$ mesa was etched out of the other and two Al strips were deposited along opposite sides on the remaining GaAs substrate to serve as excitation and detection electrodes. 
A small 
(0.3$\times$0.3~mm$^{2}$) Ohmic contact was made in the centre of this second sample, which is connected to ground (see inset in Fig. 2).
The rms excitation voltage $V$ was 10 mV. 
At low frequencies ($< 100$~kHz), the in- and out of phase components of the current $I$ induced in the detection electrode were measured 
using a current preamplifier and lock-in detector. At higher frequencies, an rf-spectrometer was used as 
receiver with a 50~$\Omega$ input impedance, much smaller than $1/\omega C$, 
where $C \sim 10^{-
12}$~F is the coupling capacitance, so that effectively the current amplitude is 
measured in this case. 
All experiments were done at a temperature of 1.5 K, where $\sigma_{xx}$ is typically $3 \times 10^{-6}~ \Omega ^{-1}$ at 5 T, falling to below $10^{-8}~ \Omega ^{-1}$ at $\nu \equiv nh/eB =2$ ($e$ elementary 
charge, $B$ magnetic field and $h$ Planck constant). Standard resistivity data 
($\rho_{xx}$ and $\rho_{xy}$) were obtained from a third conventional Hall bar 
sample from the same wafer.

A set of data for sample 1 for 4 frequencies is shown in Fig. 1. At 
the lowest frequency (10~kHz), the signal is nearly field-independent but a broad shallow ($\sim 10\%$) 
minimum around 1 T is clearly visible in addition to a shallow dip at $\nu =2$ (3.3~T).
The evolution of the pattern when the frequency is increased is shown by the next three traces, which are characterised by the persistent broad minimum, followed by a series of relatively weak (10 \%) Shubnikov-deHaas 
(SdH) oscillations and a peak, which precedes a cut-off of the signal at higher fields. The data of the traces at 53, 63 and 80 MHz are similar to those reported 
previously \cite{grodnensky,volkov}. The main peak corresponds to an Edge Magnetoplasmon (EMP) resonance which is identified by the $1/B$-
dependence of the resonant frequency \cite{volkov}. EMP's \cite{volkboek} are a manifestation 
of the Hall effect, with charge excitations localized over a distance $\ell$ near the boundary of the sample and wave velocity proportional to the Hall conductivity $\sigma_{xy} \propto 1/B$. The damping of EMP's is determined by $\sigma_{xx}$. For the 63 MHz trace, the frequency is chosen such that the EMP-resonance coincides with the 
centre of the QH-region at $\nu=2$. At the other frequencies the resonance occurs outside a QH-region, resulting in a much broader resonance.
\begin{figure}[!t]
\begin{center}
\leavevmode
\epsfig{figure=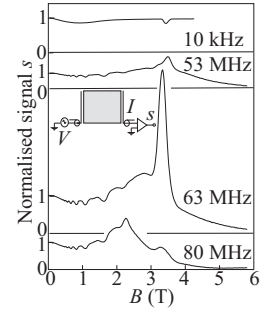,height=4cm,angle=0}
\end{center}
\caption{ The experimental capacitive response normalised to the zero-field signal, at 4 frequencies. Inset: experimental arrangement. The resonance at 63 MHz coincides with the $\nu$=2 QH-plateau.}
\label{fig1}
\end{figure}
The termination of the magnetocapacitive coupling by the $\sigma_{xy}$-controlled EMP-resonance suggests that the coupling is a result of the Hall effect. At zero field there is obviously no Hall effect, so then the coupling across the 2DES is purely resistive. As a result, it is expected that there is a gradual transition from purely resistive $\sigma_{xx}$-coupling at low fields, to Hall $\sigma_{xy}$-coupling at high fields. The resistive coupling decreases with field because of the decrease of $\sigma_{xx}$ with field, as has previously been described in the resistive plate model \cite{good}. The Hall coupling becomes more important at higher fields as then $\sigma_{xy} \gg \sigma_{xx}$. The minimum near 1 T marks the crossover between the two mechanisms. 

A direct proof for the different response mechanisms is obtained by the use of sample 2 (Fig. 2). The grounded central contact shorts out the bulk contribution, by keeping the central part of the sample at ground potential. At low frequencies, a transmission signal is only observed at the QH-regions where $\sigma_{xx} \rightarrow 0$. When the frequency 
is increased, a signal 
appears also in between the QH-regions.
At high frequencies, the transmission is 
very similar to Fig. 1, except at very low magnetic fields. This directly confirms the crossover from bulk to edge conduction. At low frequencies, the 
current through the 
central contact was also measured: see bottom trace in Fig. 2. The magnetic field dependence is 
complementary to that of the capacitive detector in the upper trace.
\begin{figure}[!t]
\begin{center}
\leavevmode
\epsfig{figure=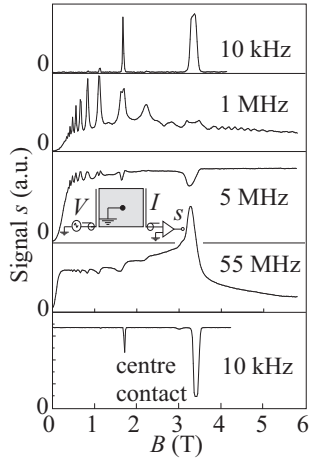,height=6cm,angle=0}
\end{center}
\caption{Upper 4 curves: the experimental capacitive response at 4 frequencies with the centre contact grounded. Inset: experimental arrangement. For this sample, the resonance at 55 MHz coincides with the $\nu$=2 QH-plateau. Lowest curve: the current through the centre contact at 10 kHz.}
\label{fig2}
\end{figure}
The capacitive response is determined by two length scales. The first is a length $\ell = \sigma_{xx}/\varepsilon_{0} \varepsilon_{r} \omega$ \cite{volkboek,noot} transverse to the edge, which corresponds to the distance over which charge penetrates the sample in a direction perpendicular to the edge in a time $1/\omega$. The second is the EMP-wavelength along the edge, given by $\lambda = \sigma_{xy}/\varepsilon_{0} \varepsilon_{r} \omega$ \cite{volkboek} (apart from a factor of order unity). Capacitive coupling occurs ({\it i}) via $\sigma_{xx}$ for $\ell \gg W$ (the sample size) and ({\it ii}) via $\sigma_{xy}$ for $\ell \ll W \ll \lambda$. A quantitative estimate for $\ell$ may be obtained from Fig. 2 for fields outside the QH-regions. The frequency of 1 MHz, which clearly marks a transition regime, together with the experimental value for $\sigma_{xx}$ near 5 T and an average $\varepsilon_{r} \simeq 5$, yields a value $\ell \sim 6$~mm, indeed in the order of the sample dimension {\it W}. Values found for $\ell$ in the domain of EMP-studies are in the order of 1~$\mu$m \cite{talyanski}. In the QH-region with $\nu =2$ ($\sigma_{xx} \ll 10^{-8}~\Omega ^{-1}$) we estimate that $\ell \ll 3$ mm at 10 kHz, consistent with the edge mode coupling. We have discussed these effects in the context of magnetocapacitance, but they are intrinsic to the sample and would also be observed for Ohmic contacts. In the domain of EMP's, the equivalence of Ohmic or capacitive contacts was recently explicitly demonstrated \cite{balaban}.

In previous works on resistive coupling \cite{good,lea}, it has been very helpful to visualise the physical concepts in terms of a series $C-R-C$ electrical circuit, with $R$ proportional to 1/$\sigma_{xx}$.
In Fig. 3, it is shown that also the basic features of the present data can be recovered by the circuit by adding an additional branch parallel to $R$ as shown in the insets in Fig 3, (c) and (d). The parallel branch consists of a (field dependent) inductance $L(B)$ to incorporate a circuit resonance (frequency $\sqrt{2/LC}$) to simulate the EMP-resonance. Guided by the theory for EMP-damping, the resistor $R_{1}$ in series with $L$ should be proportional to $\sigma_{xx}$. It should however diverge when $B \rightarrow 0$, because the Hall effect should be switched off there. The ground contact of Fig. 2 can be incorporated by splitting the resistor $R$ as in the inset to Fig. 3(d). This model is very useful in understanding the experimental response, though no universal $R(B)$ and $R_{1}(B)$ dependence will fit all the data. A typical example corresponding to the configurations of Figs. 1 and 2, is shown in Fig. 3 (c) and (d) respectively, that employ model $R(B)$ and $R_{1}(B)$ displayed in Fig. 3(b). The experimental $\sigma_{xx}$ trace is given in Fig. 3(a).
\begin{figure}[!b]
\begin{center}
\leavevmode
\epsfig{figure=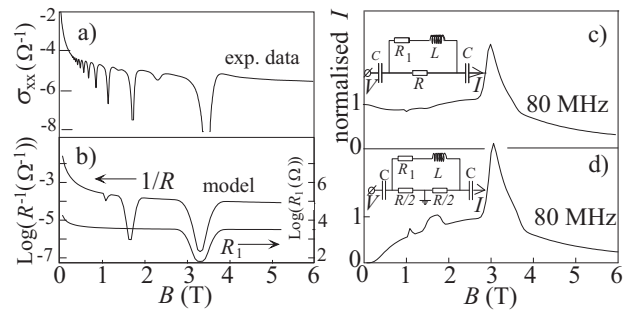,height=4cm,angle=0}
\end{center}
\caption{(a) Experimental $\sigma_{xx}$-data (converted from measured resistivities). (b) Modelled $1/R(B)$ and $R_{1}(B)$ dependencies, to generate the circuit (see insets to (c) and (d)) current $I$, normalised to zero field or with $R$=0, for (c) corresponding to the case without (comp. Fig. 1) and (d) with (comp. Fig. 2) centre contact. $L(B)$ is such that the (very broad) circuit resonance $\omega = (2/LC)^{\frac{1}{2}} $ occurs at 2.6 T.}
\label{fig3}
\end{figure}

The simulations confirm that the broad minimum near 1 T in the rf-signal comes from the crossover between bulk (i.e. $R$-branch) and edge (i.e. $R_{1}$-$L$-branch) contributions. The two configurations differ basically at low fields only, where the ground in the resistive path shorts the signal. For the experimental data, this implies that the edge contribution is well developed by 1 T at rf frequencies. Finally, at the frequency of 80 MHz, the (very broad) circuit resonance occurs at 2.6 T, outside the QH-region. Nevertheless, the peak of the signal occurs near the minimum in $\sigma_{xx}$, just as in the data.

An explicit demonstration of the existence of two coupling paths is provided by the {\em phase} behaviour of the signal at intermediate frequencies, which has a very complex field-dependence; an example is shown in Fig. 4(a).
A circuit representation is now invaluable for a qualitative understanding. Fig. 4(b) shows that the data can be nicely represented by the two-branch circuit, with a proper adjustment of the parameters (see inset). The zero-crossing of the in-phase signal basically marks the cross-over from the dominance of one branch to the other and occurs when all impedances $R$, $R_{1}$, and $1/\omega C$ are comparable.
\begin{figure}[!t]
\begin{center}
\leavevmode
\epsfig{figure=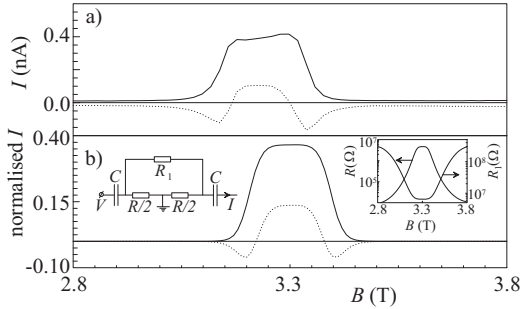,height=4cm,angle=0}
\end{center}
\caption{(a) Experimental in- and out of phase response (dashed and solid lines respectively) of the configuration with grounded centre contact at 40 kHz. (b) Normalised in- and out of phase response (dashed and solid lines respectively) of the circuit shown in the inset. Circuit parameters: $C=2 \times 10^{-12}$ F, $\omega /2 \pi =40$ kHz; $R_{1}(B)$ and $R(B)$: see inset. Note that no inductance has been used, as its impedance is negligible at low frequencies. }
\label{fig4}
\end{figure}

Away from the QH-regions at low frequencies, the characteristic length $\ell \gg W$. Consequently, as is also clear experimentally, Figs. 2 and 4, bulk resistive coupling is also important here. On the other hand, the broad minimum near 1 T in Fig. 1 was interpreted as a manifestation of the $\sigma_{xy}$ coupling mechanism. At low frequencies, outside the QH-regions, the two channels are coupled and no longer act independently. 

At sufficiently low temperatures, at the QH-regions, $\sigma_{xx}$ vanishes, and 
so would $\ell$. 
However, as is known from analysis of EMP-excitations the shape of the density 
profile near the edge is extremely important \cite{volkboek,lea,shikin}. When 
$\ell$ as defined above becomes smaller than the length $\xi$ over which the 
equilibrium charge density decays to zero at the edge (i.e. when $\omega \tau \gg 1$), $\ell$ must be replaced 
by $\xi$ \cite{volkboek,lea}. The length $\xi$ is of the order of the lateral 
depletion length (1 to 10 $\mu$m) and is (practically) independent of $\sigma_{xx}$ or 
$\sigma_{xy}$ and so of $B$. The residual capacitance at the QH-regions 
observed in the experiments of Takaoka {\em et al.} \cite{taka}, was taken as 
evidence for the existence of edge channels, whose total width is also of the 
order of the lateral depletion length \cite{shklovski}. It is important to note that such a finite residual 
capacitance is expected from an entirely classical analysis as well. The 
situation is analogous to the classical or quantum approaches of 
the high-frequency EMP \cite{talyanski}. To our knowledge a 
residual capacitance under non-QH conditions, has not yet been observed. 

From an electro-optic imaging technique employing a 5 kHz sample potential modulation a penetration length of order 100 $\mu$m was observed \cite{dietsche}. This length was associated with a surprisingly large and unexplained edge channel width. Possibly, this length might correspond to the classical length $\ell$ discussed in the present work ($\ell$ $\sim$ 100 $\mu$m for $\sigma_{xx}$ $\sim$ 10$^{-10}$ $\Omega ^{-1}$ at 5 kHz).

In conclusion, it is shown that capacitive coupling via a 2DES can occur through the Hall-effect. The coupling persists, even if $\sigma_{xx}$ is zero, provided that there is a smooth density profile near the edge, which for real systems must always be the case. The underlying principles are the same that are responsible for edge magnetoplasmons, which propagate better the smaller $\sigma_{xx}$. All observations can be understood from the local conductivities alone, without explicitly appealing to additional quantum mechanical concepts such as quantized energy levels, Quantum Hall Effect or edge channels. 

We thank W.C. van der Vleuten for sample growth and P.A.M. Nouwens for sample preparations.


\begin{references}
\bibitem{noot} For the screened case, a length $\ell ^{\prime}$ is obtained, given by $\ell ^{\prime} =\sqrt{\ell d}$, where $d$ is the distance to a screening metal plate parallel to the 2DES \cite{lea,grodnensky}.
\end{references}
\end{document}